# Label-free cell signaling pathway deconvolution of Angiotensin type 1 receptor reveals time-resolved G-protein activity and distinct AngII and AngIII\IV responses

Sandrine Lavenus [a,b,*], Élie Simard [a,b,*], Élie Besserer-Offroy [a,b], Ulrike Froehlich [a,b],

Richard Leduc [a,b,#] and Michel Grandbois [a,b,#]

**Affiliations**

[a]Department of Pharmacology-Physiology and [b]Institut de pharmacologie de Sherbrooke, Faculty of Medicine and Health Sciences, Université de Sherbrooke, Sherbrooke, Québec, Canada J1H5N4

* These authors contributed equally to this work

**email addresses**

Sandrine.Lavenus@USherbrooke.ca (SL)

Elie.Simard@USherbrooke.ca (ÉS)

Elie.Besserer-Offroy@USherbrooke.ca (ÉBO)

Ulrike.Froehlich@USherbrooke.ca (UF)

Richard.Leduc@USherbrooke.ca (RL)

Michel.Grandbois@USherbrooke.ca (MG)

**Corresponding Authors**

[#]To whom correspondence should be addressed:

Richard Leduc, Ph.D.; Richard.Leduc@USherbrooke.ca; Tel. +1 (819) 821-8000 ext. 75413

Michel Grandbois, Ph.D.; Michel.Grandbois@USherbrooke.ca; Tel. +1 (819) 821-8000 ext. 72369

Department of Pharmacology-Physiology, Faculty of Medicine and Health Sciences, Institut de pharmacologie de Sherbrooke, Université de Sherbrooke, Sherbrooke, Québec, CANADA J1H5N4

---

**Abbreviations used:** AngII, Angiotensin II; $AT_1R$, Angiotensin type 1 receptor; βarr, β-Arrestin; DMR, Dynamic mass redistribution; ERK, Extracellular signal-regulated kinase; GPCR, G protein coupled receptor; PTX, Pertussis toxin; PI3K, Phosphoinoside 3-kinase; rVSMC, rat Vascular smooth muscle cells; RVU, Reflectance variation units; ROCK, Rho-associated protein kinase; SII, [Sar[1], Ile[4], Ile[8]]AngII; SPR, Surface plasmon resonance; Src, Proto-oncogene tyrosine-protein kinase; TRV120027, [Sar[1], D-Ala[8]]AngII.

**Abstract**

Angiotensin II (AngII) type 1 receptor (AT$_1$R) is a G protein-coupled receptor known for its role in numerous physiological processes and its implication in many vascular diseases. Its functions are mediated through G protein dependent and independent signaling pathways. AT$_1$R has several endogenous peptidic agonists, all derived from angiotensinogen, as well as several synthetic ligands known to elicit biased signaling responses. Here, surface plasmon resonance (SPR) was used as a cell-based and label-free technique to quantify, in real time, the response of HEK293 cells stably expressing the human AT$_1$R. The goal was to take advantage of the integrative nature of this assay to identify specific signaling pathways in the features of the response profiles generated by numerous endogenous and synthetic ligands of AT$_1$R. First, we assessed the contributions of Gq, G12/13, Gi, Gβγ, ERK1/2 and β-arrestins pathways in the cellular responses measured by SPR where Gq, G12/Rho/ROCK together with β-arrestins and ERK1/2 were found to play significant roles. More specifically, we established a major role for G12 in the early events of the AT$_1$R-dependent response, which was followed by a robust ERK1/2 component associated to the later phase of the signal. Interestingly, endogenous AT$_1$R ligands (AngII, AngIII and AngIV) exhibited distinct responses signatures with a significant increase of the ERK1/2-like components for both AngIII and AngIV, which points toward possibly distinct physiological roles for the later. We also tested AT$_1$R biased ligands, all of which affected both the early and later events. Our results support SPR-based integrative cellular assays as a powerful approach to delineate the contribution of specific signaling pathways for a given cell response and reveal response differences associated with ligands with distinct pharmacological properties.

**Chemical compounds:** AngII (PubChem CID: 172198); AngIII (PubChem CID: 3082042); AngIV (PubChem CID: 123814); [Sar[1]]AngII (PubChem CID: 10373777); [Sar[1], Ile[8]]AngII (PubChem CID: 10079601); TRV120027 (PubChem CID: 3082475).

**Graphical Abstract**

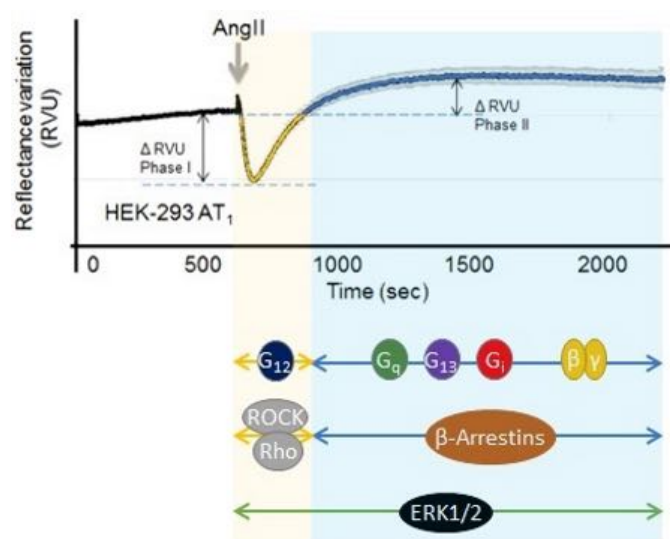

## 1. Introduction

Angiotensin II (AngII) is an octapeptide hormone that raises blood pressure through vasoconstriction and increased aldosterone biosynthesis from the adrenal cortex but also acts as a growth factor for a variety of cells such as vascular smooth muscle cells [1]. The angiotensin system contributes to the development of cardiovascular diseases [2] such as hypertension, atherosclerosis and chronic heart failure. Most of AngII's effects are caused by the activation of the angiotensin II type 1 receptor ($AT_1R$), which belongs to the G protein-coupled receptor (GPCR) superfamily [3]. Activated $AT_1R$ couples to numerous G protein-dependent signaling pathways such as the canonical $Gq$/phospholipase $C$/$IP_3$/$Ca^{2+}$/PKC pathway [3–5] and the G12/13 pathways which activate the small GTPase Rho [6–8]. Following $AT_1R$ activation, recruitment of β-arrestins (βarrs), involved in the desensitization of GPCR signaling and internalization [9–11], leads to the activation of both ERK1/2 [12] and RhoA pathways [6]. The two degradation products of AngII : AngIII (AngII (2-8)) and AngIV (AngII (3-8)), also exert their effects via $AT_1R$ [13,14]. The major physiological role of AngIII is in the brain where it is involved in the central regulation of blood pressure. It was also shown that AngIII is involved in hypertension as well as post-myocardial infarction [15,16].

In the last decade, the field of GPCR signaling has become more complex with the concept of biased signaling opening the way to the discovery of new compounds able to act as targeted therapeutics with minimal side effects. Biased agonism or functional selectivity is the ability of a ligand to promote the activation of one or several signaling pathways by favoring the receptor activation of a subset of downstream effectors. This is thought to be accomplished through the stabilization of distinct receptor conformations [17–19]. Several biased agonists of $AT_1R$ have been studied, notably [$Sar^1$, $Ile^4$, $Ile^8$]AngII (SII) which leads to $AT_1R$ phosphorylation, βarrs recruitment, receptor internalization, and βarrs-dependent ERK1/2 activation in a G protein-independent manner. This ability to antagonize $AT_1R$- dependent G protein signaling while simultaneously acting as an agonist for βarrs recruitment was such that SII became a molecule of significant therapeutic interest in cardiac pathology [20–22]. Studying $AT_1R$ bias signaling led to the development of [Sar[1], D-Ala[8]]AngII (TRV120027), a short half-life peptide that was tested in clinical trials but found to be unsuccessful for treating acute heart failure [23,24].

To fully exploit the potential of biased agonists, the exhaustive signaling profile of endogenous hormones needs to be better understood both at the molecular and the phenotypic level. Consequently, several GPCRs such as the muscarinic M2 and M3 receptors and the β2 adrenergic receptor, have been probed through the monitoring of dynamic mass redistribution (DMR), which enabled the dissection of their complex biological signaling patterns and allowed the elucidation of signaling mechanisms of newly identified GPCR-targeting compounds [25]. Similarly, to other DMR approaches based on diffraction grating, SPR measures signaling pathway activation through changes in the refractive index associated with the redistribution of the cellular molecular content present within the SPR sensor detection volume (Chabot *et al.*, 2009, 2013). Being integrative in nature, SPR monitoring integrates the ensemble of molecular events occurring as a result of receptor activation and not solely limited to one particular molecular event.

Surface Plasmon Resonance (SPR) was previously validated by our group as detection method to monitor live cells responses to various GPCRs stimulations [26,27]. In our previous work, using confocal microscopy and SPR-based imaging to monitor actin cytoskeleton remodelling and atomic force microscopy to quantify cell contraction, we have shown that early events of the $AT_1R$-dependent responses are associated with cell body contraction whereas the later phase of the response involves actin mobilization and extensive spreading of the cell [26].

Here, we used SPR, pharmacological inhibitors and siRNA to pinpoint the relative implications of Gq, G12/13, Gi, Gβγ and βarrs signaling pathways in the phenotypic responses of stimulated HEK293-AT$_1$R cells. We also tested a panel of AT$_1$R natural and biased ligands to assess their capacity to induce distinct signaling signatures. The results provide insights in the cell's phenotypic response in relationship to its signaling pathway capacities.

## 2. Material and Methods

### 2.1. Reagents

Phosphate buffered saline (PBS), trypsin/EDTA, all cell culture medium, supplements and antibiotics were purchased from Wisent (St-Bruno, Canada). UBO-QIC was purchased from Institut für Pharmazeutische Biologie, Bonn, Germany. Y-27632 was obtained from Sigma (Oakville, Canada). UO126 was purchased from Cell Signaling (Denver, USA). Gallein, pertussis toxin (PTX, 3097) and Losartan were ordered from Tocris Biosciences (Bristol, UK). All inhibitors (except PTX) were solubilized at 10 mM in pure DMSO (stock solution) and were diluted to the working concentration in assay buffer before use. PTX was dissolved in ddH$_2$O at 1 mg/mL (stock solution) and diluted in culture media before use.   AngII, AngIII, AngIV, [Sar$^1$]AngII, [Sar$^1$, Ile$^4$]AngII, [Sar$^1$, Ile$^8$]AngII, SII and TRV120027 were synthesized at our peptide synthesis facility (https://www.usherbrooke.ca/ips/fr/plateformes/psp/).

### 2.2. SPR Gold Substrate Preparation

Fisher finest premium microscope glass slides (Fisher Scientific, Ottawa, Canada) were used as base substrates. Prior to metals deposition, glass slides were cleaned in piranha solution (sulfuric acid and hydrogen peroxide 3:1) to remove any contaminants; afterwards they were placed under vacuum for metal deposition (BOC Edwards evaporator, model: AUTO 306, Crawley, UK). A chromium adhesion layer of 3 nm and a gold layer of 48 nm were deposited subsequently without breaking vacuum between evaporation. SPR substrates were coated with poly-L-lysine 0.01% (Sigma, Oakville, Canada) during 5 min to assure cell adhesion.

### 2.3. Cell Culture

HEK293 and HEK293-AT$_1$R were maintained in Dulbecco's Modified Eagle's media (DMEM) supplemented with 10 % fetal bovine serum (FBS), 2mM L-glutamine, 100IU/ml penicillin, 100µg/ml streptomycin and 0.4 mg/ml G-418 (400-130-IG). Primary rat aortic vascular smooth muscle cells (rVSMC, are a gift from Dr. Marc Servant, Université de Montréal, Montréal, QC, Canada) were maintained in DMEM supplemented with 10 % FBS, 2mM L-glutamine, 100IU/ml penicillin and 100µg/ml streptomycin. All experiments using rVSMC were conducted between passages 9 and 16. All cells lines were cultured in 100 mm petri dishes and grown at 37°C in a 5% CO$_2$ incubator. Cell culture media was refreshed every 2 days. Upon reaching confluence, cells were harvested using trypsin/EDTA for 5 min and seeded at 50,000 or 100,000 cells/cm$^2$ on either plastic petri dishes (Falcon, Reynosa, Mexico) for western blotting or poly-L-lysine coated SPR gold substrates.

## 2.4. Genetic expression knock-down using small interfering RNA (siRNA)

HEK293-AT$_1$R cells were seeded at 50,000 cells/cm$^2$ in 35mm petri dishes (Falcon, Reynosa, Mexico) and cultured overnight. Before transfection, cells were starved for 4 hours using serum-free DMEM. Two different siRNA strategies were used in order to obtain the highest efficiency. For G12/13, siRNA delivery mixes were prepared for each petri dish by adding 1.5µl of specified siRNA (20µM) in 494.5 µl of Opti-MEM I reduced serum medium (Gibco/ThermoFisher Scientific, Waltham, USA) containing 5µL of lipofectamine 2000 (Invitrogen/ThermoFisher Scientific, Waltham, USA). siRNAs delivery mixes were then incubated for 25 min at room temperature and added to the cells. After 4 hours of incubation at 37°C in a 5% CO$_2$ incubator, the cells were washed twice with antibiotic free DMEM containing 10% FBS and cultured under standard cell culture conditions until assayed. GNA12 (SASI_Hs_00150735), GNA13 (SASI_Hs_00193587) and negative control (SIC001) siRNAs were purchased from Sigma (Oakville, Canada). For βarrs, cells were transfected with siRNAs using Gene Silencer (Gene Therapy Systems, San Diego, USA) as previously described[8].

## 2.5. SPR Analysis

Cells were seeded on SPR substrates at a density of 100,000 cells/cm² until 85% confluence and starved overnight using serum-free DMEM. Cells were rinsed twice with Leibovitz's L-15 modification medium (L15) prior to the SPR experiments, which were performed at 37°C using a custom-built surface plasmon resonance apparatus described elsewhere [26,28]. In a typical SPR experiment, a 10 min baseline was established to allow cells to reach a steady state in the assay buffer (L15) or in presence of inhibitors. Afterward, AngII or other compounds diluted in assay buffer, were added and the SPR signal was monitored for 30 min. For inhibition assays, cells were pre-incubated with compounds prior AngII stimulations. All inhibitors were tested alone and were not found to elicit a SPR response. The SPR signal was acquired at 1 data point/sec. At least four independent experiments (n=4) measured in duplicate were performed for each experimental condition.

## 2.6. Western blotting

Two days after siRNA transfections, cells were washed once with cold PBS, treated for 15 min on ice with cold lysis buffer containing 1% triton and 1X protease inhibitor (Roche, Laval, Canada), then scraped and collected. Protein concentration was determined by Bradford assay (BIO-RAD, Saint-Laurent, Canada). 30μg of samples diluted in Laemmli buffer were loaded onto a 12% SDS-PAGE. Proteins were subsequently transferred onto a polyvinylidiene difluoride (PVDF) membrane (EMD Millipore, Billerica, USA). Membranes were blocked with 10% nonfat milk for an hour, then, incubated with the primary antibodies directed against GNA12 (Santa Cruz, cat. sc-409, lot G1414, Mississauga,

Canada), GNA13 (Santa Cruz, cat. sc-410, lot G0209, Mississauga, Canada), β-arrestin 1 (Cell Signaling, cat. 12697S, lot 1, Danvers, USA) or β-arrestin 2 (Cell Signaling, cat. 3857S, lot 2, Danvers, USA), overnight at 4°C. Horseradish peroxidase (HRP)-labeled Goat anti-Rabbit secondary antibody (Cell Signaling, cat. 7074S, lot 27, Danvers, USA) diluted in 10% non-fat milk and used for 1 hr at room temperature to reveal primary antibodies. Bands were detected using chemiluminescence detection substrates (BIO-RAD, Saint-Laurent, Canada). Horseradish peroxidase (HRP) labeled GAPDH (Cell Signaling, cat. 3683S, lot 4, Danvers, USA) antibody was used as loading control. Full uncropped western blot images are presented in the supplementary western blot file.

## 2.7. Data analysis

### 2.7.1. SPR analysis

Each SPR curve represents the mean of 8 SPR assays and is expressed as mean ± standard deviation (SD). Results were plotted in reflectance variation units (RVU, where 1 RVU represents 0.1% variation in total reflectance) as a function of time (sec). From those results, two parameters (namely ΔRVU phase I and ΔRVU phase II) were extracted, then transformed as % of the response observed by Ang II or Ang III, depending on the experiment, for ease of analysis. For the purpose of the phase I and II quantification, we have considered the differences between baseline just prior to AngII stimulation as shown in Fig. 1 of the manuscript.

The effect of inhibitor pre-treatment was tested by measuring the reflectance values before and after treatment (Supplementary Table S1). We have found minor variation on the basal reflectance signal after treatment with the various inhibitor. PTX data is omitted as the pre-treatment was conducted over a period of 18hrs which prevented its assessment in the current experiment configuration.

### 2.7.2. Statistical analysis

Experiments were performed at least three times in duplicate (n=3). Data are shown as mean value ± SD. In order to determine significant changes in observations, all data were analyzed using GraphPad Prism 6.05 software program (Prism, La Jolla). Briefly, the normality was tested for each set of data, and then the appropriate parametric or non-parametric test was conducted. Depending of the experiment, a two-tailed t-test for 2 groups comparison or one-way ANOVA for multiple groups comparison followed by a Dunnet's *post-hoc* test was performed.

## 3. Results

### 3.1. SPR responses of HEK293-AT$_1$R following AngII stimulation

Fig. 1 presents the integrative SPR response obtained on stable HEK293-AT$_1$R cells stimulated with 100nM AngII. Prior to AngII stimulation, a steady-state level of cellular activity was observed as indicated by the stable SPR signal. Following AngII stimulation, a rapid decline in the SPR signal (described thereafter as phase I) was observed reaching a minimum value (-22.7 ± 0.6 RVU). This transient minimum is followed by a marked increase in the SPR signal (described thereafter as phase II), which was quantified by calculating the difference between the baseline prior to stimulation and the plateau observed at 1500 s when the cell response evolved towards a later stable state (10.8 ± 1.3 RVU). As expected, AngII stimulation on Mock HEK293 cells or vehicle (L15) stimulation (Fig. 1B) on HEK 293-AT$_1$R produced no detectable SPR signal. Furthermore, the pre-incubation of HEK293-AT$_1$R with Losartan, an AT$_1$R antagonist currently used in clinic, completely abolished the AngII-mediated SPR signal (Supplementary Fig. S1), thus confirming the specificity of the SPR signal. The features (i.e. phase I and II) in the SPR response profile were then analyzed in a series of experiments designed to delineate the contribution of key components of AT$_1$R signaling with a focus on Gq, G12/13, Gi and Gβγ and βarrs.

### 3.2. Contribution of Gα$_q$ to the AngII-dependent cell response quantified by SPR

The Gq pathway is considered as a canonical component of the AngII-dependent AT$_1$R response, for it is know to produce a robust intracellular calcium mobilization [3,29]. To assess the contribution of Gq in the SPR response, we used UBO-QIC (that is, FR900359) described to specifically inhibit Gq/11/14 as observed in CHO cells [30], platelets [31] and HEK293 cells [32]. Fig. 2 shows that inhibition of Gq by UBO-QIC has a strong impact on the SPR response with an unexpected increase in phase I amplitude (169.6 ± 48.1 %) when compared to AngII stimulation alone (100.0 ± 26.9 %; Fig. 2B), as well as a robust decrease in phase II amplitude ($-78.8 \pm 75.2$ vs $100 \pm 116.2$ %; Fig. 2C). UBO-QIC was confirmed as a potent inhibitor of Gq activation by BRET (bioluminescence resonance energy transfer) biosensor analysis showing a robust decrease in IP1 production following Ang II stimulation (Supplementary Fig. S2). To assess the role of the Gq-dependent calcium component, cells were deprived in calcium using a pre-treatment with a calcium chelator (EGTA) alone or in combination with thapsigargin, a cell-permeable non-competitive inhibitor of the sarco/endoplasmic reticulum $Ca^{++}$ ATPase pump, a condition leading to the depletion of both intracellular and extracellular calcium pools. Apart from a non-significant decrease in phase II, EGTA alone or along with thapsigargin did not significantly modify the SPR response of HEK293-AT$_1$R cells to AngII. Thus, calcium mobilization normally associated with the Gq pathway does not represent a significant component of the SPR response. Interestingly, inhibition of Gq with UBO-QIC, while decreasing the phase II of the response, also produced an increased phase I, which could indicate compensatory mechanisms by other signaling pathways.

### 3.3. Contribution of Gα$_{12}$ and Gα$_{13}$ in AngII-mediated SPR cell response

Considering that the G12/13 pathways are activated by AT$_1$R and that they normally produce a robust activation of their main effector Rho/ROCK [6–8], we used Y27632, a selective ROCK inhibitor, to evaluate the contribution of these G-proteins to the AngII-mediated SPR response. Fig. 3 shows that the pre-treatment with Y27632 decreased the amplitude of phase I ($54.2 \pm 17.0$ %) when compared to AngII stimulation alone ($100.0 \pm 26.9$ %) but had no effect on the amplitude of phase II of the response. These results confirm the Rho/ROCK pathway as a major component of the AngII/AT$_1$R response measured by SPR. Since Y27632 treatment remains a rather indirect and downstream intervention to study the role of G12/13, we used siRNAs to knock-down the expression of each G protein (Fig. 3B) in HEK293-AT$_1$R cells stimulated with AngII to delineate their specific roles. When targeting G12 alone, we observed a significant decrease in the amplitude of phase I (77.4 ± 14.0 vs 119.1 ± 44.2 % for siRNA Ctl) with no significant changes in phase II as shown in Fig. 3A, C and D. In contrast, when using siRNA targeting G13 no significant change was observed in phase I while a significant decrease in phase II was observed (81.7 ± 129.6 vs 170.7 ± 96.3 % for siRNA Ctl). These results suggest distinct signaling components for both G12 and G13. Finally, the combination of siRNA targeting both G-proteins produced effects on Phase I and II consistent with treatments targeting G12 and G13 separately. These results indicate that G12/13/Rho/ROCK are important components of the HEK293-AT$_1$R SPR response to AngII, particularly in the onset of receptor activation.

### 3.4. Implication of ERK1/2 in AngII-mediated SPR response

ERK1/2, which is found downstream of multiple signaling pathways activated by AT$_1$R, is an effector expected to contribute to the global cell response measured by SPR. In presence of UO126, a MAPK/ERK kinase inhibitor, the amplitude of phase I and II (78.8 ± 15.7 and -61.1 ± 41.2 %, respectively) were found to be significantly decreased (Fig. 4), confirming the contribution of ERK signaling in the SPR response, especially in the phase II. We confirmed that UO126 was efficient at inhibiting ERK1/2 activation by observing a robust decrease in ERK phosphorylation following AngII stimulation (Supplementary Fig. S2). Next, considering the concomitant contribution of G12/Rho/ROCK, Gq and ERK1/2 pathways in the AngII-dependent response, a combination of Y27632 and UBO-QIC was used, which resulted in the abolition of the SPR cell response, thus, confirming the complementarity of these signaling pathways. Moreover, the combination of Y27632 and UO126 together also resulted in a strong decrease of the overall SPR response with notably phase I reaching 18.2 ± 3.7 % pointing toward a combined role of Rho/ROCK and ERK1/2 in the early cell response to Ang

II. Thus, ROCK and ERK1/2 activities appears to be two major drivers of the AngII-dependent response profile measured by SPR.

### 3.5. Contribution of Gi and Gβγ in AngII-mediated SPR cell response

Although some reports have described the capacity of $AT_1R$ to couple to Gi, to our knowledge coupling to Gs has not been found. Therefore, we only assessed the contributions of the Gi pathway using the inhibitor of Gi activation, PTX, an approached used previously by our group in a study of mu-opioid receptor signalling by SPR [27]. As shown in Fig. 5C, Gi inhibition significantly modified phase II of the SPR cellular response confirming the mobilization of Gi-dependent signaling. Following receptor activation and concomitant activation of G proteins, Gα dissociates from Gβγ, allowing both subunits to regulate their respective downstream signaling pathways. Gβγ signaling is complex, inhibiting or activating many downstream pathways depending on its interaction with effectors such as PI3K or Src [33,34]. To pinpoint a possible role of Gβγ-dependent signaling in the AngII SPR cellular response, we used gallein as an inhibitor of the interaction between βγ and its downstream effectors [35]. Pretreatments with gallein prior to AngII stimulation only resulted in a marginal but significant reduction of phase I amplitude ($75.0 \pm 30.6$ %) compared to the SPR response with AngII alone ($100.0 \pm 26.9$ %; Fig. 5D and E), indicating a lesser involvement of βγ-dependent signaling in the $AT_1R$-dependent SPR cellular response.

### 3.6. Contribution of β-arrestins in AngII-mediated SPR cell response

Upon activation, GPCRs are phosphorylated by GRKs leading to βarrs recruitment, desensitization of G protein-mediated signaling, receptor internalization and βarrs-dependent ERK1/2 activation. To establish the contribution of this pathway, we used siRNA against both βarrs (βarr1 and βarr2) in HEK293-AT$_1$R cells. In cells where βarr1 expression was effectively repressed (Fig. 6B) only a non-statistically significant change was observed in the phase II amplitude (Fig. 6A, C and D) when compared to control siRNA, which incidentally was found to affect the AngII-dependent response. When using siRNA targeting βarr2, a significant phase II decrease (243.0 ± 113.4 %) was observed as shown in Fig. 6A and D when compared to control siRNA (438.9 ± 246.0 %). No additional effects were found when using siRNA targeting both βarrs when compared to βarr2 alone. Altogether these results point toward a role for these pathways in the later phase of the AngII-dependent cell response.

### 3.7. Responses of rVSMC after stimulation by AngII

To assess the biological relevance of the results obtained in HEK293-AT$_1$R cells, rVSMC expressing endogenous levels of AT$_1$R were used as a more accurate physiological model. Upon stimulation of rVSMC with AngII (Fig. 7), a very similar SPR cell response was observed when compared to the HEK293-AT$_1$R cell model: a rapid decrease (phase I), followed by a gradual increase of the reflectance (phase II) that exceeds baseline values. Pre-treating cells with UBO-QIC (Fig. 7A) made no significant effect on phase I amplitude but decreased the amplitude of phase II (-92.6 ± 22.2 %) when compared to AngII alone (100.0 ± 200.8 %). Y27632 (Fig. 7A), significantly decreased both phase I and II amplitudes (42.8 ± 10.6 and -48.5 ± 26.5 %, respectively) when compared to AngII alone (100.0 ± 24.3 and 100 ± 200.8 %, respectively). These results confirmed both the implication of Gq and the Rho/ROCK pathway in the response triggered by AngII in rVSMC as observed in HEK293-AT$_1$R cells. As expected, UO126 strongly affected phase II (-120.2 ± 36.8 vs 100.0 ± 200.8 % for AngII alone), thus, also confirming an important ERK1/2 component in rVSMC. On the other hand, no significant change in the AngII response was observed following pre-treatment with gallein. Globally, this set of results demonstrates that the AngII-dependent cell SPR signature is highly similar between rVSMC and HEK293-AT$_1$R, thus validating the importance of the pathways delineated in the HEK293-AT$_1$ cell model.

### 3.8. Cellular SPR responses obtained from a panel of AT$_1$ ligands

AngIII and AngIV are two physiological products of AngII proteolysis also exhibiting agonistic activities towards AT$_1$R. Additionally, synthetic peptide analogs of AngII, namely [Sar$^1$]AngII, [Sar$^1$, Ile$^4$]AngII, [Sar$^1$, Ile$^8$]AngII and SII, were developed in order to study the structure-activity relationship of AngII with SII exhibiting AT$_1$R biased signaling properties. Since SPR monitors an integrated cell response, we hypothesized that it would be suitable for highlighting possible bias signaling subtleties on the cellular response induced by endogenous and synthetic AT$_1$R ligands. In HEK293-AT$_1$R cells stimulated with AngIII (Fig. 8A), the amplitude of phase I was found to be similar to the one induced by AngII, while the amplitude of phase II ($239.3 \pm 151.6$ %) was found to be significantly greater when compared to AngII ($100.0 \pm 116.2$ %). Stimulation with AngIV also led to similar results. These results show that distinct cell response signatures could result from AT$_1$R stimulation with the natural ligands such as AngII, AngIII and AngIV.

SPR responses were also measured for HEK293-AT$_1$R cells stimulated with different AngII analogs, some of which exhibiting biased agonist activity toward βarrs-dependent pathways (Fig. 8D and E). From the tested agonist panel, only [Sar$^1$]AngII and [Sar$^1$, Ile$^4$]AngII where found to produce a significantly lower phase I amplitude ($73.9 \pm 17.8$ and $59.3 \pm 17.8$ %, respectively) compared to AngII, whereas no significant changes in phase II were observed. With the other βarrs-biased agonists ([Sar$^1$, Ile$^8$]AngII, SII and TRV120027) no phase I response could be observed. Furthermore, only [Sar$^1$-Ile$^8$]Ang II and SII induced a small phase II-like response ($46.1 \pm 36.2$ and $31.1 \pm 48.1$ %, respectively). Taken together, these results not only indicate a limited contribution of βarrs in the AT$_1$R-dependent SPR response, but most importantly validate the sensitivity of cell-based SPR assays to pinpoint differences in the responses generated by natural ligands.

### 3.9. Implication of Gα$_q$, Gα$_{12/13}$ and Gβγ in AngIII SPR cell response

Given the difference observed between the SPR profiles of AngII and AngIII, we performed additional experiments to identify the signaling components leading to distinct AngIII-dependent cellular response. Our data on AngII demonstrated that Gq, G12/13/ROCK, Gβγ and ERK1/2 were involved to various degrees in the SPR signal of AngII. UBO-QIC pre-treatment (Fig. 9A, B and C) did not significantly modify the amplitude of phase I induced by AngIII but strongly decreased the amplitude of phase II (-27.0 ± 9.2 %) when compared to AngIII alone (100.0 ± 63.4 %). In consistency with the observations made for AngII, this result shows a major implication of the Gq pathway in the phase II of the AngIII-dependent SPR response profile. Y27632 was also used to evaluate the contribution of the Rho/ROCK pathway in the AngIII cellular response (Fig. 9A, B and C). A decrease in the amplitude of the phase I cell response to AngIII (38.5 ± 15.7 %) was observed, a result also consistent with the one obtained previously with AngII. As expected, ERK1/2 inhibitor UO126 strongly decreased the amplitude of phase II (-21.9 ± 20.3 %) when compared to AngIII alone, confirming that an important ERK signaling component is also mediated by AngIII. Finally, pre-treatment of cells with gallein resulted in a decreased amplitude of phase II (41.6 ± 37.4 %) when compared to AngIII alone, indicating that βγ signaling is a likely component of the ERK1/2 response following the activation of the Gq pathway.

## 4. Discussion

In our study, the main features of the SPR response profile of cells expressing AT$_1$R were analyzed in a series of experiments designed to delineate the contribution of key components of signaling with a focus on Gq, G12/13, Gi and Gβγ and βarrs. We found that following AngII stimulation of HEK293-AT$_1$R cells, a rapid decline in the SPR signal (described previously and thereafter as phase I) was observed. This transient minimum was followed by a marked increase in the SPR signal (described previously and thereafter as phase II). Therefore, for ease of analysis in our study, the SPR signal was treated as 2 distinct phases, based on our previous findings where we showed that phase I of AT$_1$R-dependent cell responses in HEK293-AT$_1$R cells and VSMC were attributed to cell body contraction whereas phase II was associated with actin mobilization and extensive spreading of the cell [26].

Gq/11 signaling has long been described as a primary transduction mechanism initiated by AngII in neuronal, cardiac and smooth muscle cells [3]. Accordingly, pre-treatments with the Gq/11 inhibitor UBO-QIC led to a robust decrease of the late events (phase II) associated with the AngII-dependent response of HEK293-AT$_1$R cells (Table 1) and in rVSMC (Table 3). Interestingly, the inhibition of calcium signaling with EGTA alone or in combination with the thapsigargin did not produce any significant effect on the early AngII response (Table 1), suggesting that calcium mobilization *per se* does not lead to intracellular changes detectable by SPR. These results also confirmed previous studies in which the morphological response of HEK293-AT$_1$R cells stimulated with AngII was found to be largely independent from intracellular Ca$^{2+}$ mobilisation [26].

G12 and G13 are well known to activate small GTPases of the Rho family. These proteins are key regulators of the cytoskeleton dynamic involved in cell contraction and motility [36–38] and are activated by AngII in cells expressing AT$_1$R [6–8]. Using Y27632, a ROCK inhibitor, we confirmed a robust contribution of the Rho/ROCK pathway in both HEK293-AT$_1$R cells (Table 1) and rVSMC (Table 3). The specific contributions of both G12 and G13 were also assessed in this study using siRNAs. We observed a role for G12 in the early events (phase I) which support recent data by our group using BRET biosensors [39,40] and from other research groups [41]. We also observed a role of G13 in the later phase of the signal (Table 2), where its relative weight in the response observed must be interpreted cautiously given the slight effects of the transfection procedure on the general response profile of AngII. Taken together, these results, point toward a specific and robust contribution of the G12/Rho/ROCK pathway in the early cellular events following $AT_1R$ stimulation in both HEK293-$AT_1R$ cells and rVSMC.

Interestingly, Gq inhibition by UBO-QIC amplified phase I while ROCK inhibition by Y27632 significantly decreased phase I (Table 1). One might have predicted here that the effects of UBO-QIC and Y27632 used together could have cancelled out each other. Instead, the combination of Y27632 and UBO-QIC (Table 1) resulted in the blockade of the SPR cell response suggesting a compensatory effect between the Gq and Rho/ROCK pathways. Thus, the inhibition of Gq by UBO-QIC could favor G12/G13/Rho coupling and associated cell contraction (i.e. phase I). These findings are in line with the G12/13-mediated mechanical response of HEK293-$AT_1R$ after stimulation by AngII monitored by atomic force microscopy (AFM) [42] as well as the contraction of rVSMC observed after treatment by Endothelin 1 (ET-1). Endothelin receptor ($ET_A$) is another GPCR well known to trigger a G12/13 response [43]. Our findings are also in line with previous *in vitro* observations suggesting a direct role for ROCK on the status of the acto-myosin contractile machinery [44]. Indeed ROCK was found to directly phosphorylate and activate the myosin light chain (MLC) *in vivo* [45,46] and to inactivate the myosin light chain phosphatase (MLCP) thus promoting acto-myosin contractile activity [47]. Interestingly, ROCK-inhibition was also was also found to inhibit contraction *in vivo* suggesting a central role in the regulation of the activity of the contractile machinery [48,49]. Despite all these evidences for an integrated role of $Ca^{2+}$ and Rho/ROCK pathways, additional investigations are required to better understand the intricate interactions between elements of the G12/G13/Rho and Gq pathways.

It was previously shown in rat adrenal glomeruli, liver, kidney, and pituitary cells that $AT_1R$ couples to Gi/o proteins, thus inhibiting adenylyl cyclase activity [3,50]. Gi is also the only signaling pathway activated by $AT_1R$ after ligand-independent mechanoactivation [51]. Our results with HEK293-$AT_1R$ cells treated with PTX, a Gi/o inhibitor [52], revealed a contribution of these G proteins in the phase II of the cellular response measured by SPR (Table 1) that could be due to a modulation of ERK1/2 activity [3,52]. Indeed, among the other effectors activated by $AT_1R$, ERK1/2 activation is triggered via several intracellular signaling pathways either that are G protein-dependent [4,5] or independent, such as βarrs [12]. The use of UO126, a MEK1/2 inhibitor, showed a major implication of ERK1/2 signaling in the phase II of the cellular response for both the HEK293-$AT_1R$ (Table 1) and rVSMC (Table 3) cell models. Interestingly, we found that ERK1/2 activation was mainly dependent upon Gq/11 signaling as ERK1/2 activation assays (data not shown) showed that UBO-QIC treatment almost abolished AngII-induced ERK1/2 activity. Co-inhibition of ERK1/2 and ROCK signaling confirmed both effectors as the two major contributions to the cellular response (Table 1).

The βγ G protein subunits are also known to directly modulate the activities of many effectors [34]. For instance, βγ dimers can recruit PI3Kγ to the plasma membrane, resulting in increased ERK1/2 activity. It is widely accepted that the activation of G proteins by $AT_1R$ also induces βγ signaling associated with tyrosine kinases activation [53,54]. Pre-treatment of HEK293-$AT_1R$ cells with gallein, a βγ inhibitor [35], indicated a marginal role of βγ signaling in the early events following $AT_1R$ activation (Table 1).

$AT_1R$ recruits βarrs [9–11], which drives myosin light chain phosphorylation and cell contraction as demonstrated by force measurement on individual cells [8]. βarrs were also shown to produce a delayed and sustained activation of ERK1/2 pathway [10,55], thus suggesting a contribution of these proteins in the integrated cell response. Indeed, the small but significant phase II-like response obtained with βarrs-biased ligands [Sar$^1$-Ile$^8$]AngII and SII (Table 4) is consistent with a contribution of this pathway in the cellular response following AT$_1$R activation. The specific contribution of βarr1 and βarr2, was assessed using siRNA and results showed that βarr2 contributes to the phase II of the AT$_1$R-dependent SPR response in HEK293-AT$_1$R cells (Table 2), consistent with the activation of ERK1/2 pathway downstream of βarrs [55]. Again here, the extent of their role in the response observed must be interpreted cautiously given the observed impact of the transfection procedure on the cells.

The first part of our study has allowed us to delineate the contribution of key signaling pathways in the global cell response induced by the natural AT$_1$R ligand, AngII. This knowledge was next used to differentiate the cell responses induced by a panel of endogenous and synthetic AT$_1$R ligands. Even if AngII and AngIII have been described to activate the same signaling pathways notably Gq, G12, βarr1 and 2 as well as a robust downstream activation of ERK1/2 [39], we observed that both ligands produced distinct SPR response profiles whereas, the AngIV response (Table 4) remained mainly similar to the one observed with AngII. Interestingly, the response to Ang III produced a phase II (Table 4) of large amplitude compared to AngII, suggesting enhanced ERK1/2 activity. In support to this assertion, phase II was indeed prevented in cells pre-treated either with UBO-QIC or UO126 in a similar fashion for both AngIII (Table 5) and AngII (Table 1), thus confirming the major role of ERK1/2 in that phase. Although AngII, AngIII and AngIV activate the same signaling pathways [39] our results suggest that all three ligands are distinct in terms of their efficacy in mediating downstream ERK1/2 effects, which could lead to distinct physiological outcomes for these ligands.

In previous studies, several AngII analogs were used in order to perform structure-activity relationship study of AT$_1$R. These studies suggested that position 1 of AngII has very little, if no impact at all, on $AT_1R$ functional selectivity [56], G12 activation [39], inositol phosphate production [20] or ERK1/2 activation [57]. For this reason, [Sar[1]]AngII, (Table 4) has been described as an $AT_1R$ agonist similar to AngII [39]. However, the results presented herein show a small but significant difference in the amplitude of phase I when comparing the response profiles of [Sar[1]]AngII and AngII (Table 4). Examining the SPR response obtained with [Sar[1]-Ile[4]]AngII, an analog described as less efficient to signal via Gq and βarrs [39], we observed a lower phase I amplitude when compared to AngII, consistent with a lesser G protein contribution for this ligand. Three other biased agonists, namely [Sar[1]-Ile[8]]AngII, SII and TRV120027 have been described as βarrs-biased when compared to G protein signaling [12,58,59]. As expected, these ligands induced a marginal SPR cell response for both phase I and II, thus indicating a rather small contribution of βarrs to the specific signaling patterns we observed in HEK293-$AT_1R$ cells.

When using cellular response signature based on SPR to identify specific signaling pathways involved in $AT_1R$ signaling and to characterize the pharmacological properties of ligands, one must be careful when applying findings to a biological context. Indeed, it is known that different cellular models, physiological or not, often respond differently to identical treatments, regardless of the output measured, as the nature and context in terms of signaling elements of each cell type can affect their phenotypic response. In our study, HEK293-$AT_1R$ cells were used as an easily workable model especially regarding transfection efficacy. Interestingly, HEK293-$AT_1R$ and VSMC responses to $AT_1R$ activation measured by SPR were found to be similar, which support the use of HEK293-$AT_1R$ as a valid model for investigating $AT_1R$ signaling *in vitro*.

## 5. Conclusion

Cell-based SPR assays allow the characterization of integrated and time-resolved cellular responses, which involves multiple signaling pathways, therefore providing new basis for future SPR studies focused on determining the cellular response to signalling bias. Such studies could allow both the dissection of complex signaling patterns of GPCRs in basic research and the understanding of mechanisms of drug action in the context of GPCR drug discovery. More specifically, using SPR as a label-free cellular assay, we demonstrated that HEK293-AT$_1$R and rVSMC exhibit very similar response profile [12,58,59] confirming HEK293-AT$_1$R as a valid model for the study of AT$_1$R signaling pathways related to vascular pathophysiology. Our assays also allowed us to determine and discriminate the response profile of a panel of ligands, which led to the observation of a striking difference between cellular responses as seen in the initial phase, which was found to be strongly dependent on the G12/Rho/ROCK pathways, and the second phase, associated with a robust ERK1/2 contribution mainly dependent on Gq but seemingly also involving Gi/G13 and associated Gβγ and βarrs signaling activity. While most AT$_1$R pathways previously identified by other groups were also observed in this study, we gained deeper insights on their relative weight within a time-resolved integrated cell response. Finally, we observed striking differences between the cell responses to endogenous ligands AngII and AngIII/AngIV, mainly attributed to a greater ERK1/2 component suggesting different biological effects.

## Acknowledgements

SL was the recipient of a Post-doctoral Fellowship from the FRQ-S-funded Centre de recherche du Centre hospitalier universitaire de Sherbrooke. ÉBO was supported by a research fellowship from the Institut de pharmacologie de Sherbrooke (IPS).

### *Funding*

This work was supported by the National Science and Engineering Research Council of Canada (NSERC) to MG and the Canadian Institute of Health Research (CIHR) to RL. The authors declare no competing financial interests.

### *Authors contribution*

Conception and design of study: SL, ÉS, ÉBO, MG, and RL

Acquisition of data: SL, ÉS, ÉBO, and UF

Analysis and interpretation of data: SL, ÉS, ÉBO, UF, MG, and RL

Wrote the manuscript: SL, ÉS, ÉBO, MG, and RL

All the authors approved the version of the manuscript to be published.

**Figures and Tables**

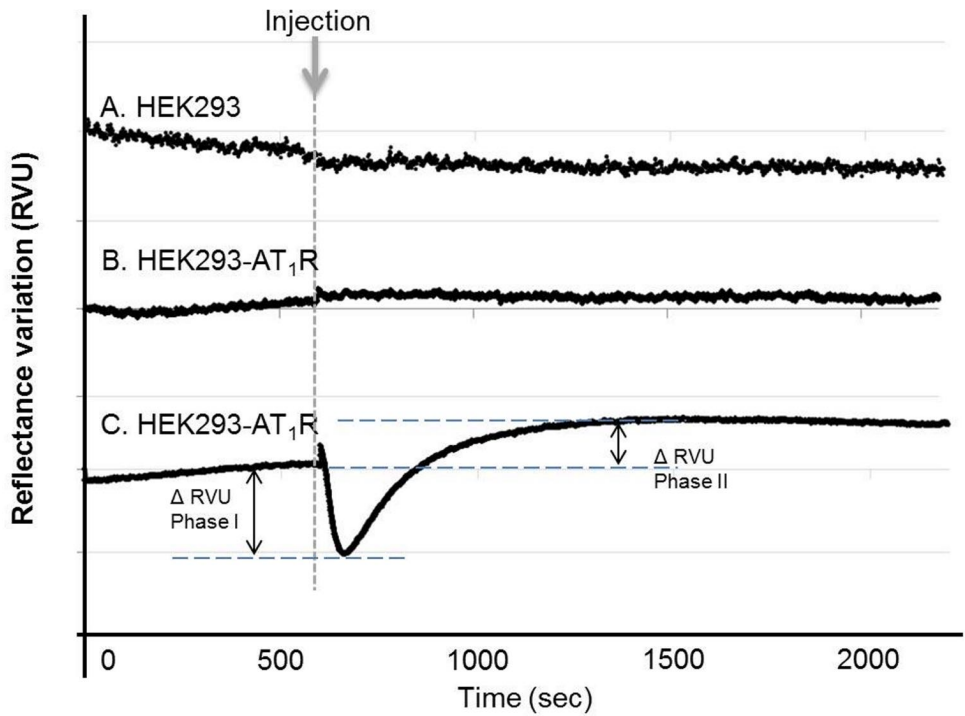

**Figure 1. SPR responses of HEK293-AT$_1$R stimulated by AngII.**

Reflectance variations as function of time of Mock cells stimulated with 100 nM AngII (**A**) and HEK293-AT$_1$R stimulated with Vehicle (**B**) or 100 nM AngII (**C**). Injection was done at 600 s (10 min) as indicated by the gray arrow and dashed line. Each curve represents the mean of 8 independent experiments. Distance between Y axis graduation (gray line) correspond to 0.1 RVU.

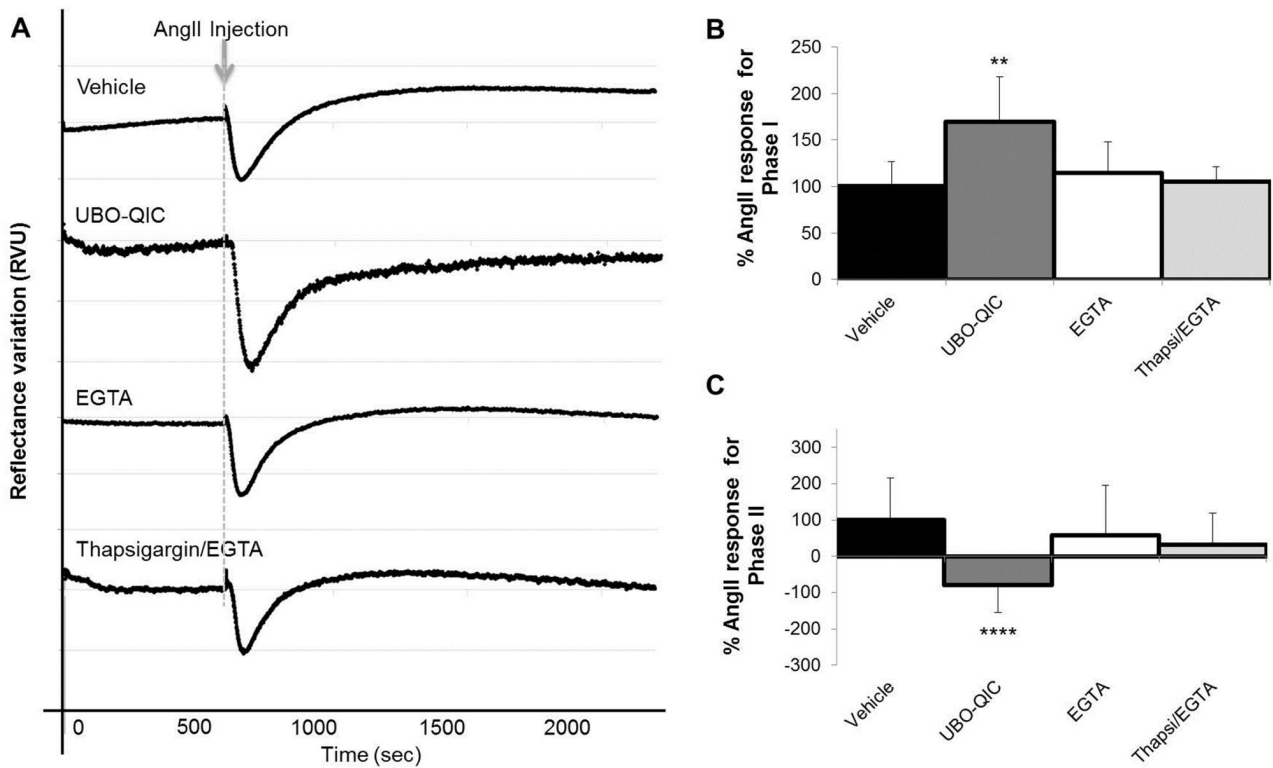

**Figure 2. Contribution of Gq in the SPR response induced by AngII on AT$_1$R.**

(**A**) Reflectance variations as function of time of HEK293-AT$_1$R cells stimulated with 100 nM AngII and pre-treated with Vehicle, UBO-QIC (1 μM, 10min, Gq inhibitor) or Thapsigargin (1 μM, 10min, inhibitor of the sarco/endoplasmic reticulum Ca$^{2+}$ATPase). AngII injections are indicated by the gray arrow and dashed line. 2mM EGTA was added to assay medium during the EGTA and Thapsigargin experiments. Distance between Y axis graduation (gray line) correspond to 0.1 RVU. (**B**) Amplitudes of the reflection variations for phase I and (**C**) amplitudes of the reflection variations for phase II expressed as % of the AngII response. Each curve and bar represents the mean of 8 independent experiments and are expressed as mean ± SD. ** $p < 0.01$ and *** $p < 0.001$ compared to vehicle.

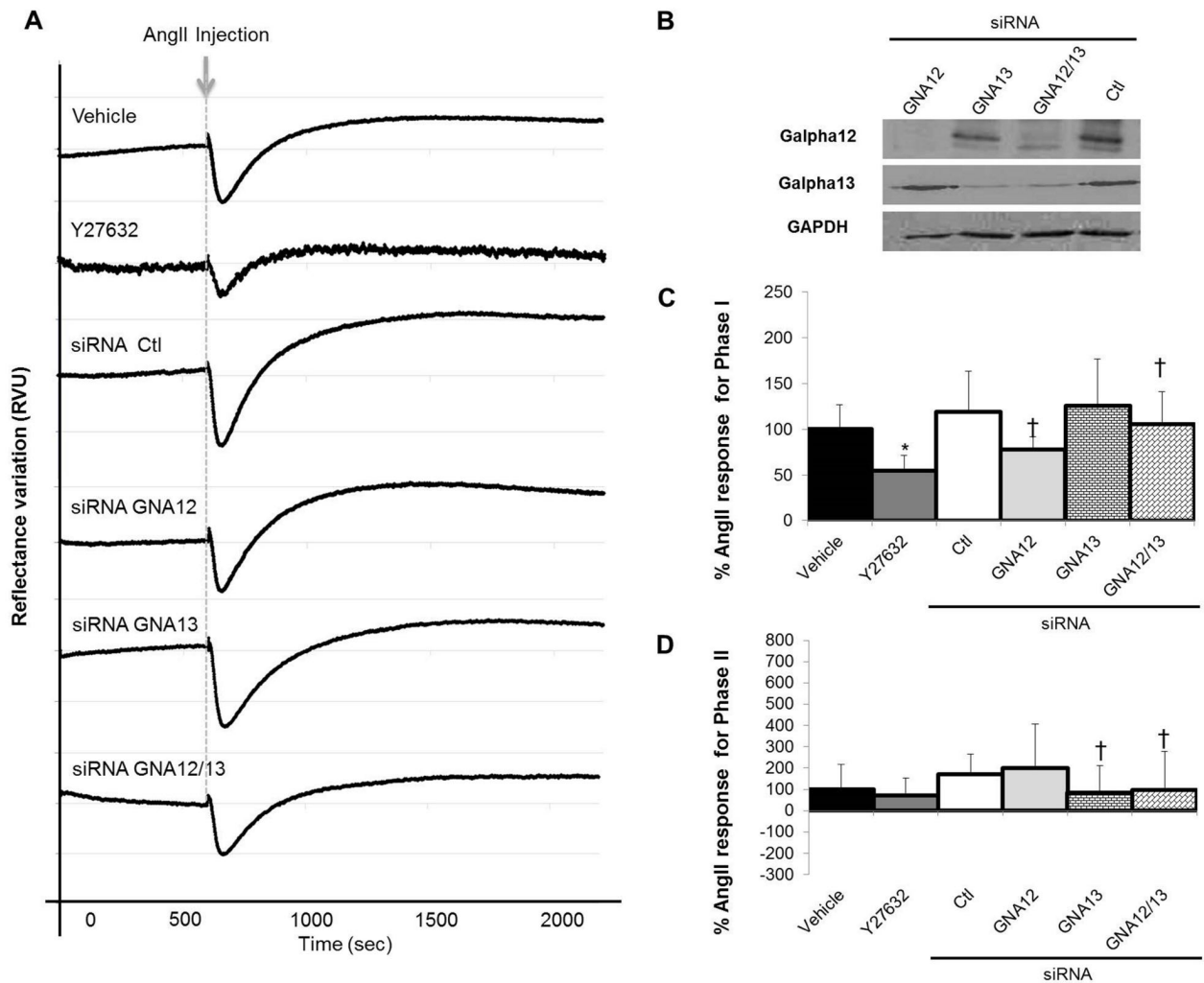

**Figure 3. Contribution of G12/13 in the SPR response induced by AngII on AT$_1$R.**

(**A**) Reflectance variations as function of time of HEK293-AT$_1$R stimulated with 100 nM AngII and pre-treated with Vehicle, Y27632 (10μM, 10 min, ROCK inhibitor) or transfected with Ctl siRNA, G12 siRNA, G13 siRNA or G12 and G13 siRNAs. AngII injection is indicated by the gray arrow and dashed line. Distance between Y axis graduation (gray line) correspond to 0.1 RVU. (**B**) Expression of G12 and G13 after siRNA transfection measured by immunoblotting. HEK293-AT$_1$R cells were transfected with Ctl siRNA or siRNAs targeting G12 and or G13 mRNA. The expression of GAPDH under each condition was used as loading control. Amplitudes of the reflection variations for phase I (**C**) and phase II (**D**) expressed as % of the AngII response. Each curve and bar represents the mean of 8 independent experiments and are expressed as mean $\pm$ SD. $\dagger$ $p < 0.05$ compared to siRNA Ctl and * $p < 0.05$ compared to vehicle.

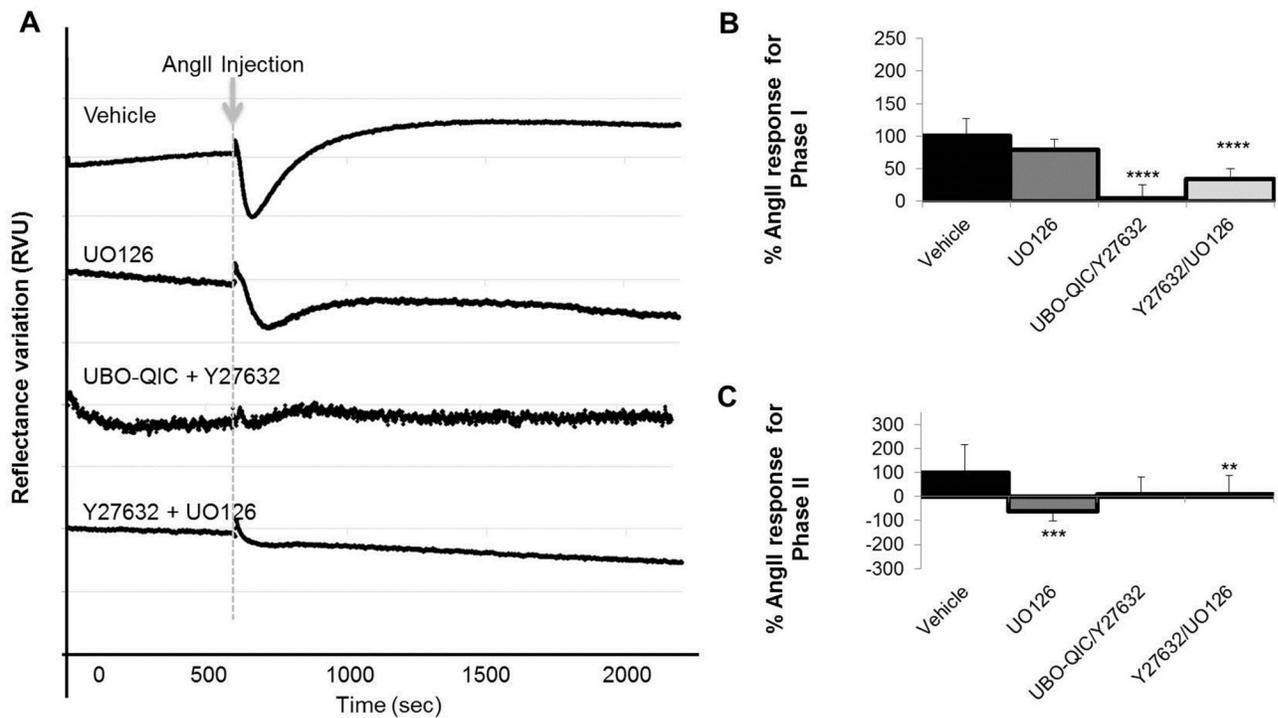

**Figure 4. Implication of multiple pathways activation in the AngII SPR response.**

(**A**) Reflectance variations as function of time of HEK293-AT$_1$R stimulated with 100 nM AngII and pre-treated with Vehicle, UO126 (10μM, 30min, MEK inhibitor), UBO-QIC/Y27632 (1μM/10μM, 10min), or Y27632/UO126 (10μM/10μM, 30min). AngII injection is indicated by the gray arrow and dashed line. Distance between Y axis graduation (gray line) correspond to 0.1 RVU. Amplitudes of the reflection variations for phase I (**B**) and phase II (**C**) expressed as % of the AngII response. Each curve and bar represents the mean of 8 independent experiments and are expressed as mean ± SD. ** $p < 0.01$, *** $p < 0.001$, and **** $p < 0.0001$ compared to vehicle.

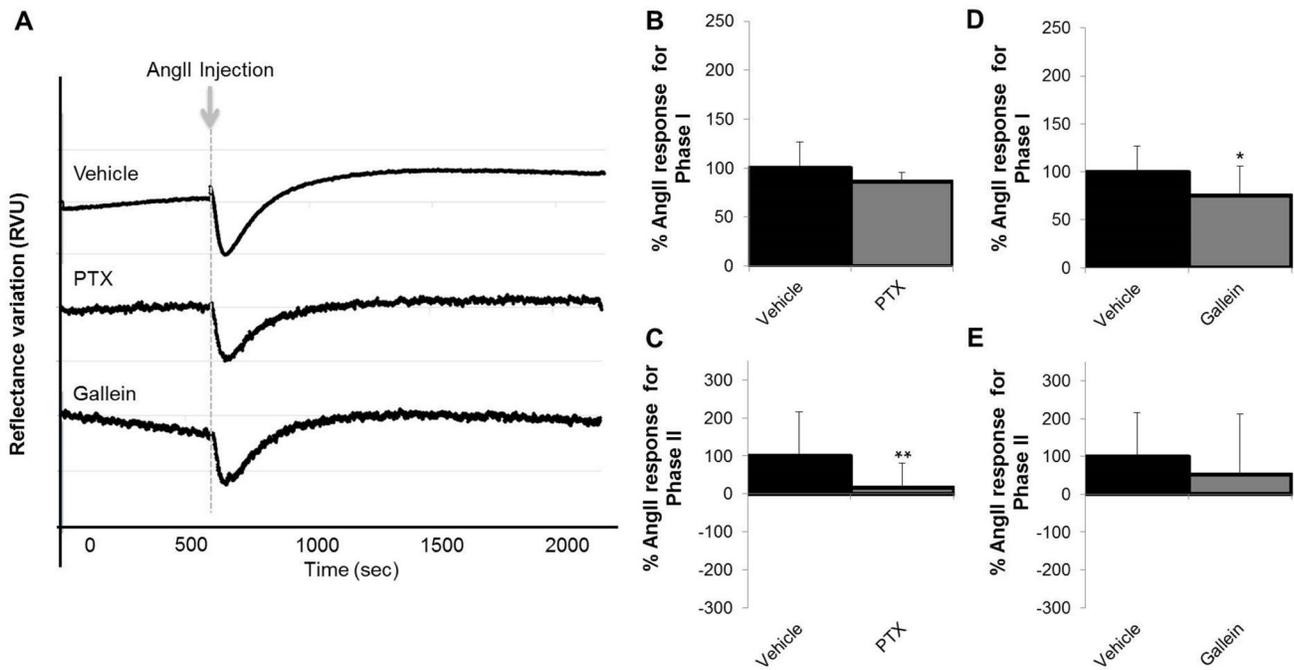

**Figure 5. Contributions of Gi and Gβγ in the SPR response induced by AngII on AT$_1$R.**

(**A**) Reflectance variations as function of time of HEK293-AT$_1$R stimulated with 100 nM AngII and pre-treated with Vehicle, PTX (100ng/ml, overnight) or Gallein (20μM, 30min, βγ signaling inhibitor). Distance between Y axis graduation (gray line) correspond to 0.1 RVU. Amplitudes of the reflection variations for phase I (**B** and **D**) and phase II (**C** and **E**) expressed as % of the AngII response. Each curve and bar represents the mean of 8 independent experiments and are expressed as mean ± SD. * $p < 0.05$ and ** $p < 0.01$ compared to vehicle.

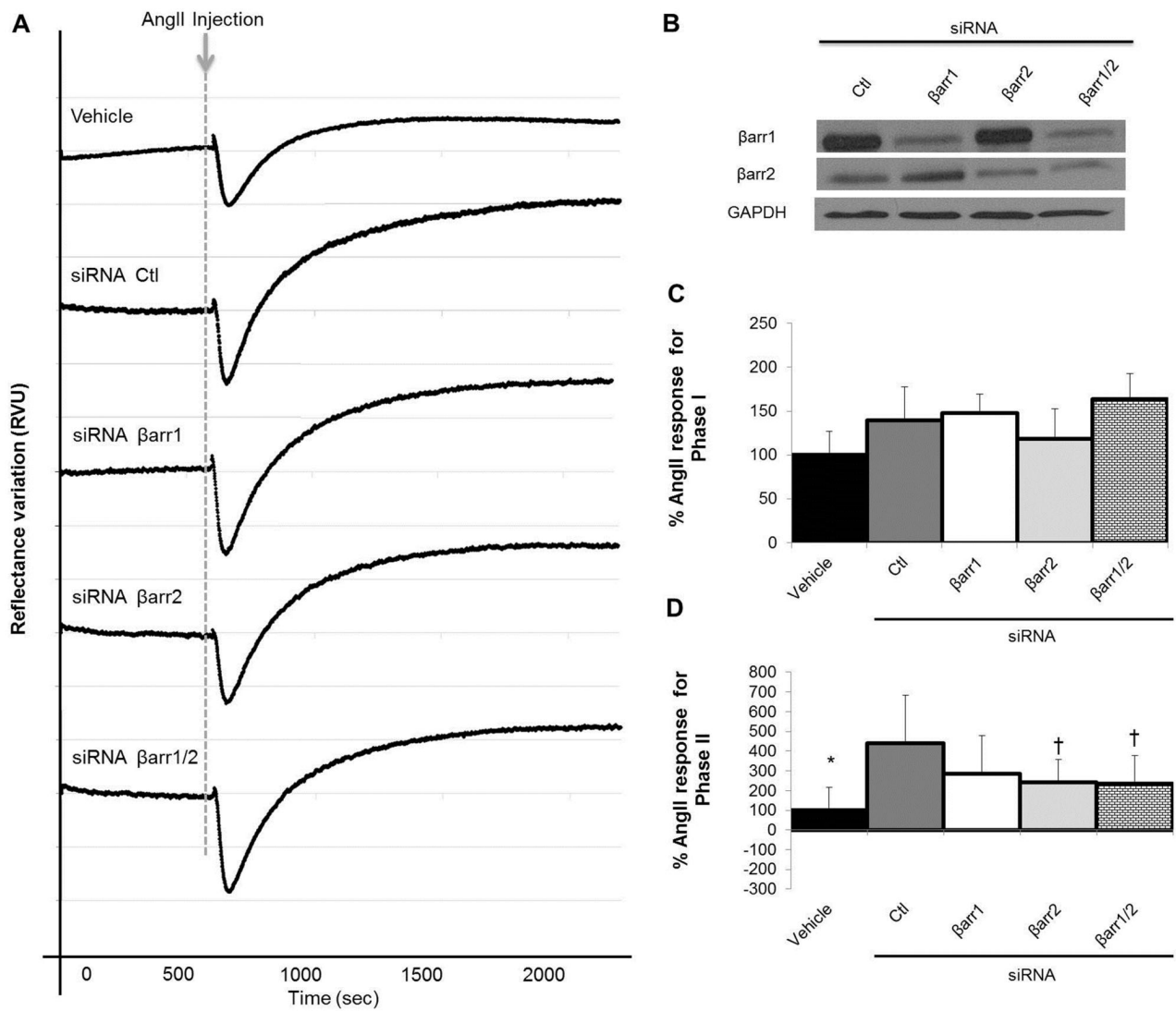

**Figure 6. Contribution of βarr1 and βarr2 in the SPR response induced by AngII on AT$_1$R.**

(**A**) Reflectance variations as function of time of HEK293-AT$_1$R stimulated with 100 nM AngII and pre-treated with Vehicle, or transfected with Ctl, βarr1, βarr2 or βarr1 and 2 siRNAs. AngII injection is indicated by the gray arrow and dashed line. Distance between Y axis graduation (gray line) correspond to 0.1 RVU. (**B**) Expression of βarr1 and 2 after siRNA transfection measured by immunoblotting. HEK293-AT$_1$R cells were transfected with Ctl siRNA or siRNA targeting βarr1 and/or βarr2 mRNA. The expression of GAPDH under each condition was used as loading control. Amplitudes of the reflection variations for phase I (**C**) and phase II (**D**) expressed as % of the AngII response. Each curve and bar represents the mean of 8 independent experiments and are expressed as mean ± SD. † $p < 0.05$ compared to siRNA Ctl and * $p < 0.05$ compared to vehicle.

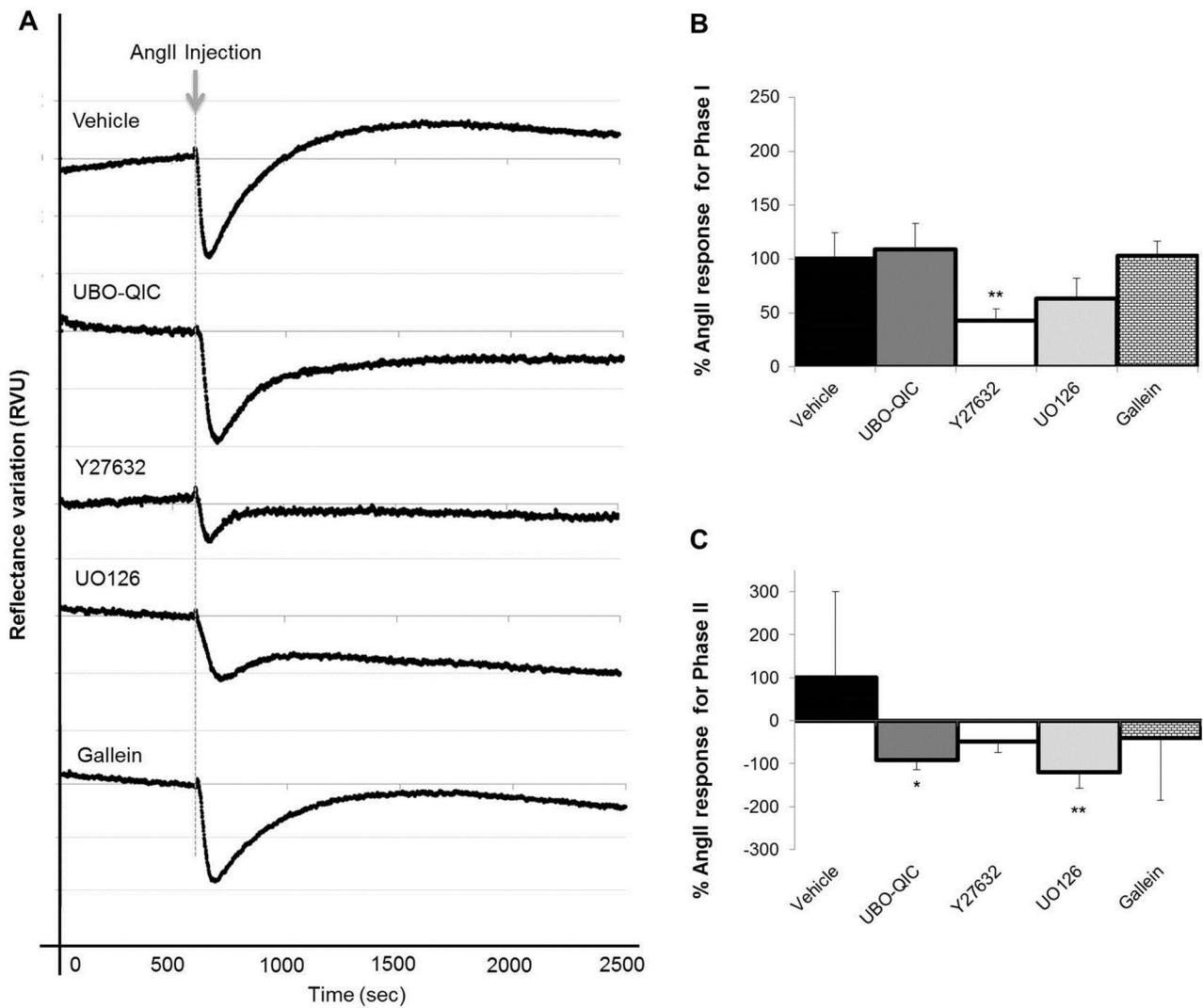

**Figure 7. SPR responses of rVSMC stimulated by AngII.**

(**A**) Reflectance variations as function of time of rVSMC stimulated with 100 nM AngII and pre-treated with Vehicle, UBO-QIC (1µM, 10min), Y27632 (10µM, 30min), Gallein (20µM, 30min) or UO126 (10µM, 30min) respectively. AngII injection is indicated by the gray arrow and dashed line. Distance between Y axis graduation (gray line) correspond to 0.1 RVU. Amplitudes of the reflection variations for phase I (**B**) and phase II (**C**) expressed as % of the AngII response. Each curve and bar represents the mean of 4-6 experiments and are expressed as mean ± SD. * $p < 0.05$ and ** $p < 0.01$ compared to vehicle.

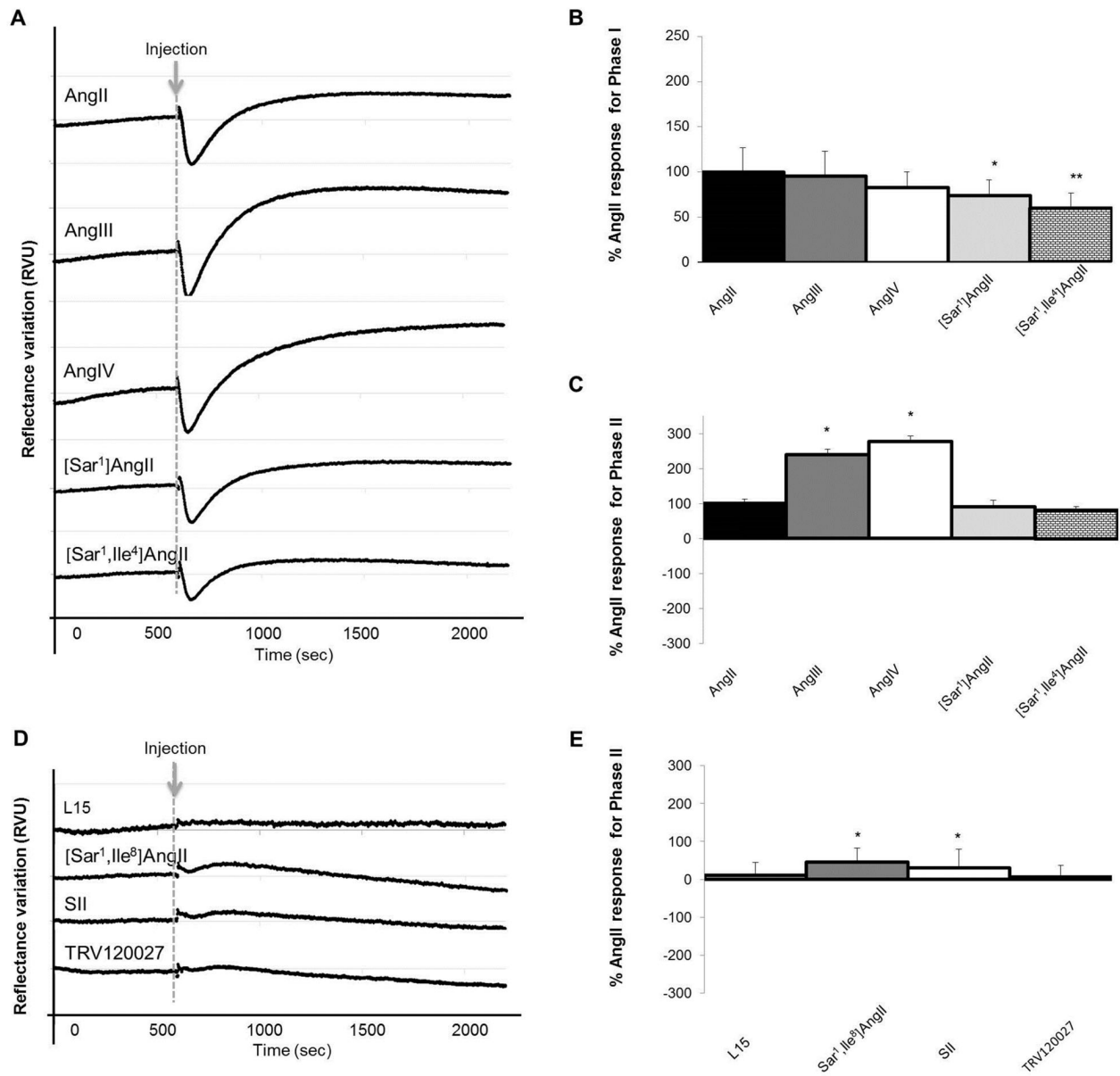

**Figure 8. SPR responses associated with other AT₁R ligands.**

(**A**) Reflectance variations as function of time of HEK293-AT₁R after injection of 100nM AngII, 1μM AngIII or 10μM AngIV. Distance between Y axis graduation (gray line) correspond to 0.1 RVU. Amplitudes of the reflection variations for phase I (**B**) and phase II (**C**) expressed as % of the AngII response. (**D**) Reflectance variations as function of time of HEK293-AT₁R after injection of 100nM

[Sar$^1$]AngII, 10µM [Sar$^1$, Ile$^4$]AngII, 1µM [Sar$^1$, Ile$^8$]AngII, 10µM SII or 100nM TRV120027. Distance between Y axis graduation (gray line) correspond to 0.1 RVU. Amplitudes of the reflection variations for phase II (**E**) expressed as % of the vehicle response. Each curve and bar represents the mean of 8 independent experiments and are expressed as mean ± SD. * $p < 0.05$ and ** $p < 0.01$ compared to vehicle. Due to the affinity difference toward AT$_1$R, the final concentration of each ligand was adjusted to reach the same receptor occupancy, i.e. approximatively 100 times Ki, which implies working at 100nM for AngII, 1µM for AngIII, 10µM for AngIV, 100nM for [Sar$^1$]AngII, 10µM for [Sar$^1$, Ile$^4$]AngII, 1µM for [Sar1, Ile$^8$]AngII, 10µM for SII and 100nM for TRV120027. Injection is indicated by the gray arrow and dashed line.

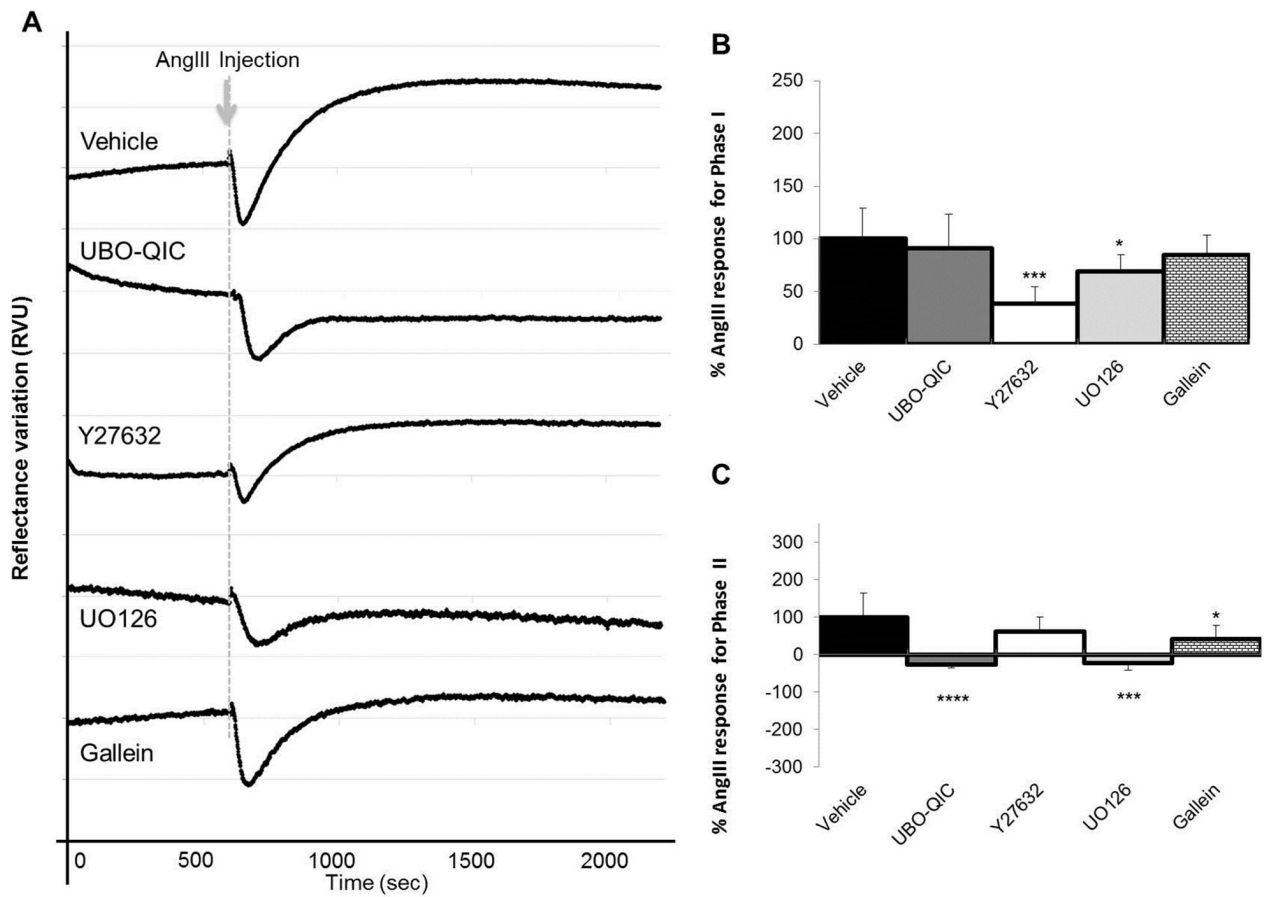

**Figure 9. Implication of the Gq, G12/13 and Gβγ in the AngIII SPR response.**

(**A**) Reflectance variations as function of time of HEK293-AT$_1$R stimulated with 1 μM AngIII and pre-treated with Vehicle, UBO-QIC (1μM, 10min), UO126 (10μM, 30min), Y27632 (10μM, 30min) or Gallein (20μM, 30min) respectively. AngIII injection is indicated by the gray arrow and dashed line. Distance between Y axis graduation (gray line) correspond to 0.1 RVU. Amplitudes of the reflection variation for phase I (**B**) and phase II (**C**) expressed as % of the AngIII response. Each curve and bar represents the mean of 8 independent experiments and are expressed as mean ± SD. * $p < 0.05$, *** $p < 0.001$, and **** $p < 0.0001$ compared to AngII or AngIII alone).

**Table 1. Effects of all inhibitors on Ang II-induced SPR responses in HEK293-AT$_1$R cells.**

Amplitudes of the reflection variations for phase I and for phase II expressed as % of the AngII response.

Each data represents the mean of 8 independent experiments and is expressed as mean ± SD. * $p < 0.05$,

** $p < 0.01$ and *** $p < 0.001$ compared to vehicle.

| Pre-treatment | Pathway | Phase I | Phase II |
|---|---|---|---|
| Vehicle | - | 100.0 ± 26.9 | 100 ± 116.2 |
| UBO-QIC | Gq | 169.6 ± 48.1** | -78.8 ± 75.2** |
| EGTA | Calcium$^a$ | 114.2 ± 33.8 | 57.4 ± 137.6 |
| Thapsigagin + EGTA | Calcium$^b$ | 105.4 ± 15.5 | 32.3 ± 87.7 |
| Y27632 | Rho/ROCK | 54.2 ± 17.0* | 71.0 ± 83.2 |
| UO126 | MEK/ERK1/2 | 78.8 ± 15.7* | -61.1 ± 41.2* |
| UBO-QIC + Y26632 | Gq + Rho/ROCK | 3.9 ± 21.4*** | 10.4 ± 70.3 |
| Y27632 + UO126 | Rho/ROCK + MEK/ERK1/2 | 18.2 ± 3.7*** | -43.8 ± 34.4** |
| PTX | Gi | 86.3 ± 9.5 | 15.8 ± 65.2* |
| Gallein | Bγ | 75.0 ± 30.6* | 51.9 ± 160.5 |

$^a$ Condition leading to depletion of extracellular calcium.

$^b$ Condition leading to depletion of both intracellular and extracellular calcium.

**Table 2. Effects of siRNAs treatments on Ang II-induced SPR responses in HEK293-AT$_1$R cells.**

Amplitudes of the reflection variations for phase I and for phase II expressed as % of the AngII response.

Each data represents the mean of 8 independent experiments and is expressed as mean ± SD. * $p < 0.05$ compared to Lipofectamine, and † $p < 0.05$ compared to Gene Silencer.

| Pre-treatment | Pathway | Phase I | Phase II |
|---|---|---|---|
| Vehicle | - | 100.0 ± 26.9 | 100.0 ± 116.2 |
| Lipofectamine | - | 119.1 ± 44.2 | 170.7 ± 96.3 |
| siRNA GNA12 | G12 | 77.4 ± 14.0* | 200.9 ± 206.6 |
| siRNA GNA13 | G13 | 126.0 ± 50.8 | 81.7 ± 129.6* |
| siRNAs GNA12 + GNA13 | G12 + G13 | 105.3 ± 35.5* | 94.6 ± 182.9* |
| Gene Silencer | - | 139.9 ± 37.9 | 438.9 ± 246.0 |
| siRNA ARRB1 | βarr1 | 148.1 ± 21.2 | 285.5 ± 194.3 |
| siRNA ARRB2 | βarr2 | 118.1 ± 34.9 | 243.0 ± 113.4† |
| siRNAs ARRB1 + ARRB2 | βarr1 + βarr2 | 163.7 ± 28.7 | 232.7 ± 145.6† |

**Table 3. Effects of all inhibitors on Ang II-induced SPR responses in rVSMC.** Amplitudes of the reflection variations for phase I and for phase II expressed as % of the AngII response. Each data represents the mean of 8 independent experiments and is expressed as mean ± SD. * $p < 0.05$ and ** $p < 0.01$ compared to vehicle.

| Pre-treatment | Pathway | Phase I | Phase II |
|:---:|:---:|:---:|:---:|
| Vehicle | - | 100.0 ± 24.3 | 100.0 ± 200.8 |
| UBO-QIC | Gq | 108.6 ± 24.6 | -92.6 ± 22.2* |
| Y27632 | Rho/ROCK | 42.8 ± 10.6** | - 48.5 ± 26.5 |
| UO126 | MEK/ERK1/2 | 63.4 ± 18.5 | -120.2 ± 36.8** |
| Gallein | βγ | 102.5 ± 13.9 | -40.3 ± 144.4 |

**Table 4. SPR responses of multiple known AT$_1$R ligands in HEK293-AT$_1$R cells.** Amplitudes of the reflection variations for phase I and for phase II expressed as % of the AngII response for AngII to [Sar$^1$,Ile$^4$]AngII and % of the Vehicle response for Vehicle to TRV120027. Each data represents the mean of 8 independent experiments and is expressed as mean ± SD. * $p < 0.05$ and ** $p < 0.01$ compared to AngII, and † $p < 0.05$ compared to vehicle.

| Treatment | Phase I | Phase II |
|---|---|---|
| AngII | 100.0 ± 26.9 | 100.0 ± 116.2 |
| AngIII | 95.5 ± 27.6 | 239.3 ± 151.6* |
| AngIV | 82.5 ± 17.8 | 276.8 ± 156.8** |
| [Sar$^1$]AngII | 73.9 ± 17.8* | 91.1 ± 167.7 |
| [Sar$^1$,Ile$^4$]AngII | 59.3 ± 17.8** | 82.0 ± 90.1 |
| L15 | -4.4 ± 7.0 | 11.1 ± 33.5 |
| [Sar$^1$,Ile$^8$]AngII | 12.1 ± 9.9 | 46.1 ± 36.2† |
| SII | -4.4 ± 9.7 | 31.1 ± 48.1† |
| TRV120027 | -0.63 ± 12.3 | 6.6 ± 31.4 |

**Table 5. Effects of all inhibitors on Ang III-induced SPR responses in HEK293-AT$_1$R cells.**

Amplitudes of the reflection variations for phase I and for phase II expressed as % of the AngIII response.

Each data represents the mean of 8 independent experiments and is expressed as mean ± SD. * $p < 0.05$,

** $p < 0.01$ and *** $p < 0.001$ compared to vehicle.

| Pre-treatment | Pathway | Phase I | Phase II |
|:---:|:---:|:---:|:---:|
| Vehicle | - | 100.0 ± 28.9 | 100.0 ± 63.4 |
| UBO-QIC | Gq | 90.9 ± 32.8 | -27.0 ± 9.2*** |
| Y27632 | Rho/ROCK | 38.5 ± 15.7*** | 61.7 ± 38.3 |
| UO126 | MEK/ERK1/2 | 69.3 ± 15.8* | -21.9 ± 20.3** |
| Gallein | βγ | 84.2 ± 19.4 | 41.6 ± 37.4* |

**Label-free cell signaling pathway deconvolution of Angiotensin type 1**

**receptor reveals time-resolved G-protein activity and distinct**

**AngII and AngIII\IV responses**

## _Supporting Information_

Sandrine Lavenus[a,b,*], Élie Simard [a,b,*], Élie Besserer-Offroy [a,b], Ulrike Froehlich [a,b],

Richard Leduc [a,b,#] and Michel Grandbois [a,b,#]

**Affiliations**

[a]Department of Pharmacology-Physiology and [b]Institut de pharmacologie de Sherbrooke, Faculty

of Medicine and Health Sciences, Université de Sherbrooke, Sherbrooke, Québec, Canada J1H5N4

* These authors contributed equally to this work

[#]To whom correspondence should be addressed

**TABLE OF CONTENT**

**Supplementary Materials and Methods**

*BRET² Assay*

HEK293 were transiently co-transfected with plasmids containing cDNA encoding for hAT$_1$-GFP10 and RlucII-G$\alpha_q$. The next day, transfected cells were detached and seeded at a density of 50,000 cells/well into white 96-wells plates (Falcon). 48 hours post-transfection, cells were washed once with PBS then 90µL of HBSS containing 20mM HEPES was added. Prior stimulation with ligands, cells were pre-treated with either Vehicle (L15) or UBO-QIC (1µM) for 10min. Then, Coelenterazine-400A (5µM, obtained from Gold Biotechnology Inc.) was added following by ligands at increasing concentrations. BRET measurements were measured using a TECAN M1000 fluorescence plate reader and the BRET ratio was calculated by dividing the GFP10 emission by the luminescence emission.

*IP-One Assays*

IP-One assay was performed according to the manufacturer recommendations. Briefly, HEK293-AT$_1$ cells (15 000 cells/wells, 384 shallow-well plate) were pre-treated with Vehicle (L15) or UBO-QIC (1µM) for 10min, then an increasing concentrations of AngII was added for 30 min. At the end of the incubation time, IP1-d2 and anti-IP1-Cryptate were added for at least 1 h. Plates were then read on a GENios Pro plate reader with HTRF filters (excitation at 320 nm, emission at 620 and 665 nm). TR-FRET ratio was determined as the fluorescence of the acceptor (665 nm) over the fluorescence of the donor (620 nm).

*AlphaScreen assay*

ERK1/2 phosphorylation in HEK293-AT$_1$ cells was monitored using Perkin Elmer's AlphaScreen SureFire assays (PerkinElmer, Billerica, MA) according to the manufacturer recommendations. Briefly, cells were seeded in 96-well plates at 50 000 cells/well.

The next day, cells were starved overnight. After 48 hours, cells were pre-treated with vehicle (L15), UBO-QIC (1µM, 10min), Y27632 (10µM, 10min), Gallein (20µM, 30min) and UO126 (10µM, 30min), then stimulated with 100nM AngII for 0, 3, 5, 10, 15, 30 and 60 min. At the end of the incubation time, cells were lysed with 25 µL of 5X lysis buffer and incubated at room temperature for 10 min on a plate shaker and then frozen overnight at -20 °C. 5 µL of the lysate was used for analysis. Readings were performed on a PerkinElmer EnSpire 2300 multilabel reader

**Supplementary Table S1. Effect of inhibitor treatment on SPR signal.**

| INHIBITOR | ABSOLUTE REFLECTANCE (RVU ± SD) | | |
|---|---|---|---|
| | Before inhibitor treatment | Before AngII injection | Variation (%) |
| **GALLEIN** | 0.603 ± 0.024 | 0.580 ± 0.022 | - 3.81 |
| **UBO-QIC** | 0.649 ± 0.030 | 0.645 ± 0.026 | 0.61 |
| **UO126** | 0.596 ± 0.009 | 0.583 ± 0.005 | - 2.18 |
| **Y27632** | 0.626 ± 0.019 | 0.626 ± 0.023 | 0.00 |

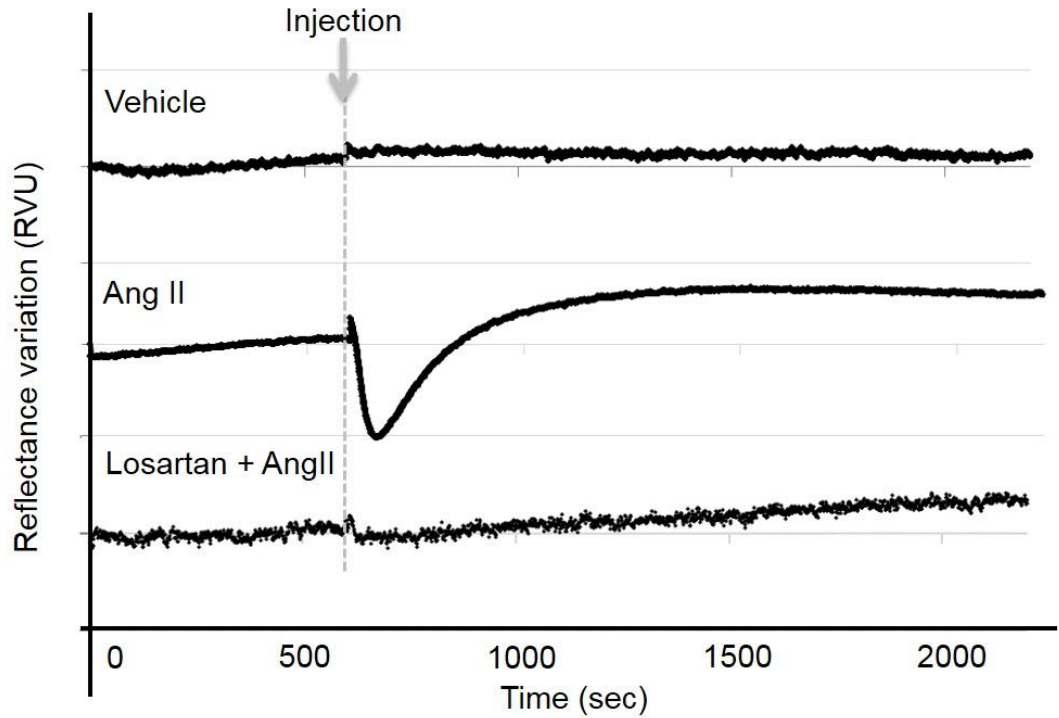

**Supplementary Fig. S1. AngII-induced SPR signal is mediated through AT₁R.**

Reflectance variations as function of time of HEK293-AT$_1$R after injection of vehicle, 100nM

AngII or 100 nM AngII after 10 min pre-treatment with 10 µM Losartan. Distance between Y

axis graduation (gray line) correspond to 0.1 RVU.

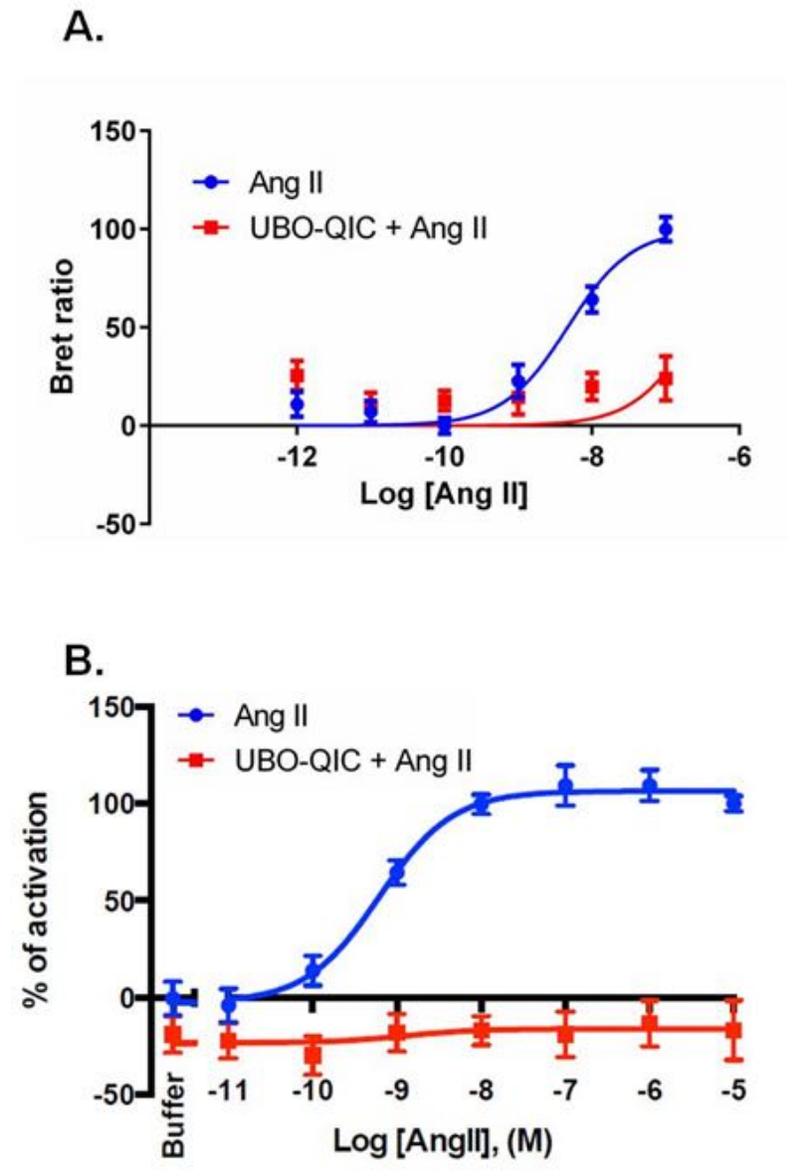

**Supplementary Fig. S2. UBO-QIC effect on Gq.**

(**A**) Gαq activation in HEK293-AT₁R stimulated with increased concentrations of AngII and pre-treated with Vehicle or UBO-QIC (1μM, 10min). (**B**) Inositol 1-phosphate production in HEK293-AT₁R stimulated with increased concentrations of AngII and pre-treated with Vehicle or UBO-QIC (1μM, 10min).

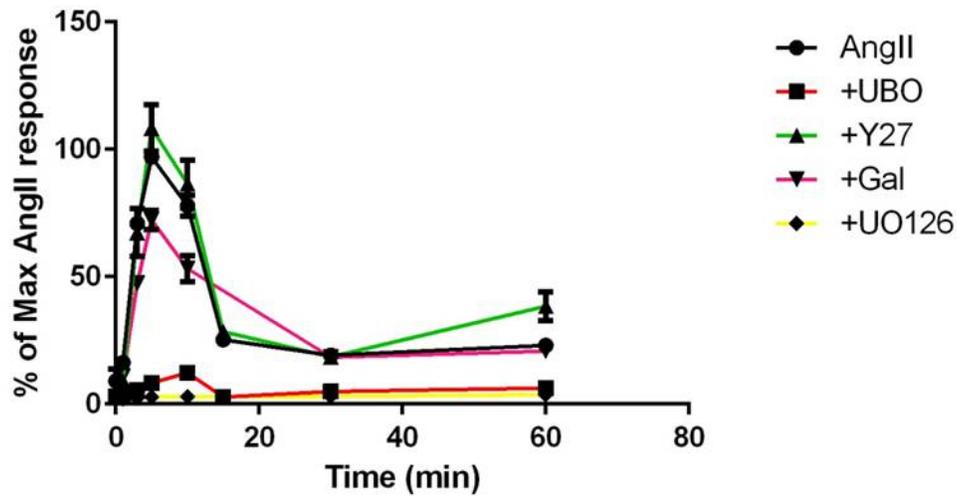

**Supplementary Fig. S3. Activation of ERK1/2 following AngII stimulation.**

ERK1/2 phosphorylation kinetics in HEK293-AT$_1$R stimulated with 100nM of AngII and pre-treated with vehicle, UBO-QIC (1µM, 10min), Y27632 (10µM, 10min), Gallein (20µM, 30min) and UO126 (10µM, 30min). The data are normalized to the AngII maximum effect. Each curve represents the mean ± SEM of at least three independent experiments performed in duplicate.

# Label-free cell signaling pathway deconvolution of Angiotensin type 1 receptor reveals time-resolved G-protein activity and distinct agonist responses

## _Supplementary Western Blots_

Sandrine Lavenus[a,b,*], Élie Simard [a,b,*], Élie Besserer-Offroy [a,b],

Richard Leduc [a,b,#] and Michel Grandbois [a,b,#]

**Affiliations**

[a]Department of Pharmacology-Physiology and [b]Institut de pharmacologie de Sherbrooke, Faculty of Medicine and Health Sciences, Université de Sherbrooke, Sherbrooke, Québec, Canada J1H5N4

* These authors contributed equally to this work

[#]To whom correspondence should be addressed

**TABLE OF CONTENT**

## Supplementary Western Blots Images S1

### *siRNA GNA12/13*

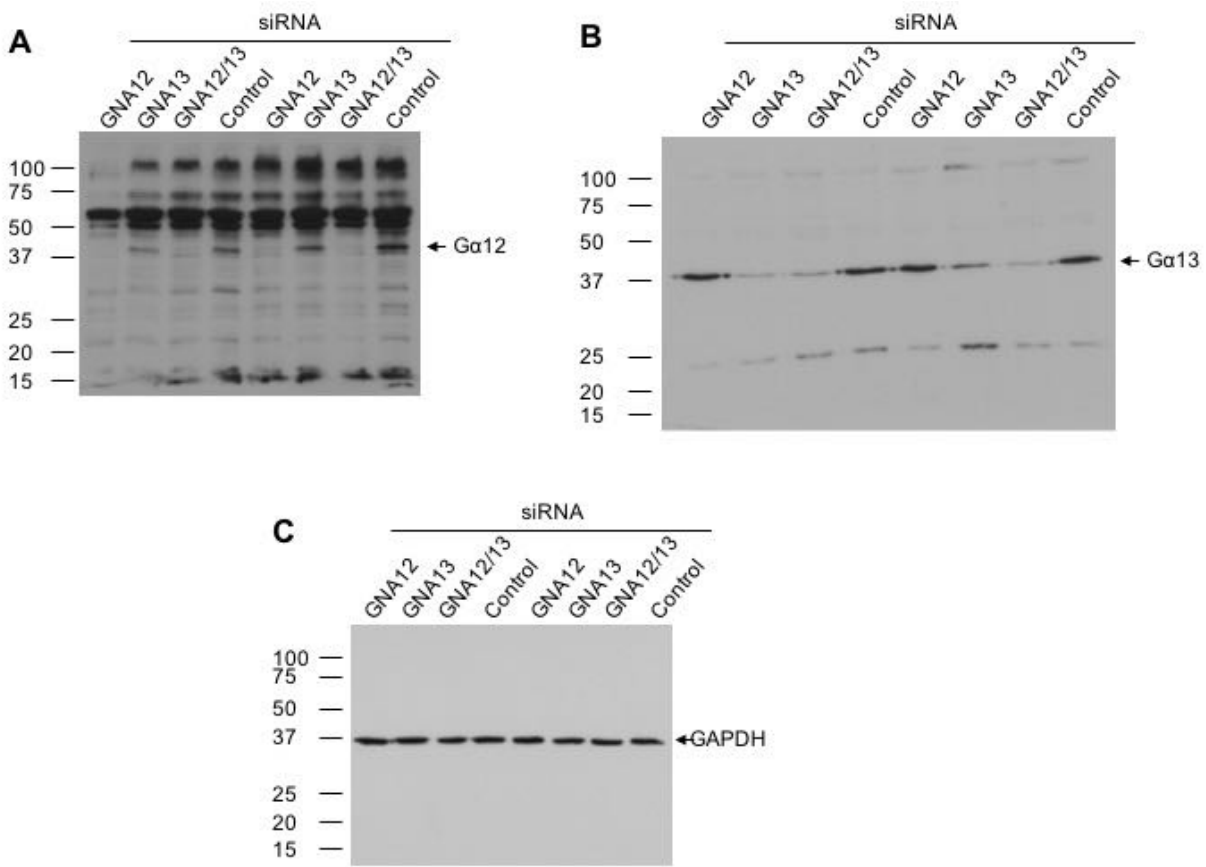

**Supplementary Western Blots S1.** Uncropped images of the western blots for the knock-down validation of Gα12 and Gα13. (**A**) Revelation of the Gα12 immunoreactivity, (**B**) Revelation of the Gα13 immunoreactivity, and (**C**) Revelation of the GAPDH immunoreactivity.

**Supplementary Western Blots Image S2**

*siRNA ARRB1/2*

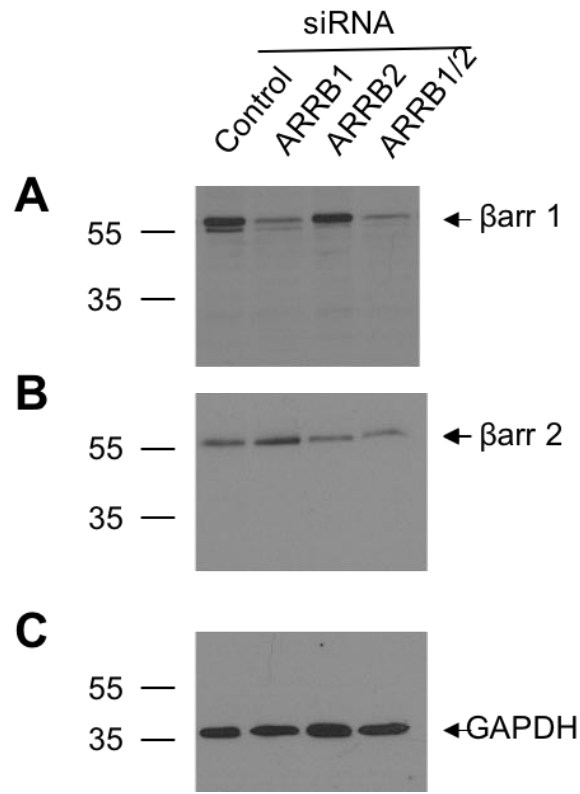

**Supplementary Western Blots S2.** Uncropped images of the western blots for the knock-down validation of βarr 1 and βarr 2 . (**A**) Revelation of the βarr 1 immunoreactivity, (**B**) Revelation of the βarr 2 immunoreactivity, and (**C**) Revelation of the GAPDH immunoreactivity.